\documentstyle[12pt,epsf]{article}

\def\balpha{\mbox{\boldmath$\alpha$}}

\def\bsigma{\mbox{\boldmath$\sigma$}}
\def\bgamma{\mbox{\boldmath$\gamma$}}

\def\sD{{\tiny D}}
\def\sL{{\tiny \lambda}}
\def\sW{{\tiny W}}

\def\b1{{\bf 1}}

\def\bp{{\bf p}}
\def\br{{\bf r}}
\def\bv{{\bf v}}

\def\bx{{\bf x}}
\def\by{{\bf y}}
\def\bs{{\bf s}}

\def\bw{{\bf w}}
\def\bL{{\bf L}}

\def\bD{{\bf D}}
\def\bB{{\bf B}}

\def\bE{{\bf E}}

\newcommand{\be}{\begin{equation}}
\newcommand{\ee}{\end{equation}}
\newcommand{\bea}{\begin{eqnarray}}
\newcommand{\eea}{\end{eqnarray}}

\newcommand{\lan}{\langle}
\newcommand{\ran}{\rangle}

\begin{document}

\vskip 2cm
\begin{center}
\Large 
{\bf  Revisiting the Eichten - Feinberg - } \\
{\bf   Gromes $ Q \bar{Q}$  Spin-Orbit Interaction } \\
\vskip 0.5cm
\large
 Ken Williams \\
{\small \em Department of Physical and Earth Sciences,  } \\
{\small \em Jacksonville State University, Jacksonville, 
Alabama 36265 } \\
\end{center}
\thispagestyle{empty}
\vskip 0.7cm

\begin{abstract} 

  Invariant and covariant forms of the quark-antiquark interaction
  derived by the method of Eichten and Feinberg are considered.
  Relations between the various terms imposed by Lorentz
  transformation constraints, here called Gromes relations, are found
  to exist in neither case. Details of the Gromes relation proper are
  reconsidered and inconsistencies found that lead to a violation of
  covariance.
\end{abstract}

\newpage 

\section{Introduction} 

Constituent quark models provide valuable insight into many hadronic
phenomena - the mass spectrum as well as decay and transition
observables. With the appearance of advanced probes such as CEBAF,
where high resolution data on fine and hyperfine spectra to the order
of 200 - 300 kev. are expected with complimentary data on hadronic
production, their predictions are of increasing interest and in
increasing use. In both sectors non-perturbative QCD effects are
expected to be significant.

Theoretical input for modeling the non-perturbative regime has come
from the strong coupling limit of lattice gauge theory. Wilson loop
area asymptotics \cite{wilson} for heavy static constituents lead
directly to the popular linear confinement model. When quarks are
given motion their Wilson loop and the potential interaction derived
from it are then manifestly non-local, embodying the confining
non-perturbative gluon dynamics. A proper accounting for these effects
to $O(m^{-2}) $ has been made in the well-known work of Eichten and
Feinburg (E.F.), ref\cite{ef}, where they enter implicitly. They do
not however enter into reductions of relativistic equations that input
only the linearly confining potential; there the semi-relativistic
corrections are purely kinematical. Hence an increased interest in the
E.F. spin dependent interaction and the simplifying relation of Gromes
\cite{gr}.

It would be difficult to overstate the impact these two results have
had on particle physics calculations since their appearance and
general acceptance over a decade ago ( Spires preprint database alone,
e.g., lists a combined $\approx 700 $ citation entries.); it extends
from phenomenological modeling to lattice calculations. The aim of the
present article is to help clarify their meanings and to set forth
criticisms in the most accessible terms available. The results are
reviewed separately beginning in the next section with that of
ref\cite{ef} where it is demonstrated that whether in its invariant or
covariant form there are with the potential no accompanying Gromes
relations(G.R.). In the section that follows the G.R. proper is
treated by way of its derivation and is found to violate Lorentz
covariance. The violation is in consequence of two main errors in the
derivation: 1) discrepancy between the Lorentz transformation and the
one employed there as such, and, 2) imposing of Lorentz invariance
where covariance is required.
 
\section{ E.F. spin-orbit potential}

Beginning from the gauge invariant quark-antiquark transition
amplitude, a linearly confining potential for static quarks is defined
in terms of the Wilson loop
\bea
\lefteqn{ G_{I,c} = \langle 0| T^* \bar{\psi}^c(y_2)
P(y_2, y_1) \psi(y_1) \bar{\psi}(x_1) P(x_1, x_2) \psi^c(x_2) |0
\rangle } \label{one} \\ & & \to \langle tr_\sD tr_\sL \, \, S_0(y_1, x_1 ;A) P(x_1, x_2) C^{-1} S_0(x_2, y_2 ;A) C P(y_2, y_1)\rangle
\label{two} \\ && = - tr_\sD \left(\frac{1+\gamma^0}{2} \otimes \frac{1-\gamma^0}{2} \, \tilde{I} \right) \, e^{-i(m_1+m_2) T} \delta(\bx_1-\by_1) \delta(\bx_2-\by_2) \, , \\
\lefteqn{ \epsilon(r)\equiv -\frac{1}{T} \ln(\tilde{I})
= -\frac{1}{T} \ln\langle tr P \exp(\imath g
\oint dz_\mu A^\mu(z)) \rangle \equiv -\frac{1}{T} \ln \lan 1 \ran_\sW } \label{three}
\eea
for large $ T \equiv x^0-y^0 $. The symbols above have the following meanings:
$T^* $ time orders, $ C $ charge conjugates, the $ Ps $
are path-ordered exponentials, $ P(x,y) = P \exp[\imath g \int_y^x
dz_\mu A^\mu(z) ] $, $ tr_\sD ( tr_\sL ) $ is the trace operator for Dirac (gauge) matrices, and the average is taken over the gauge fields, $ \langle U
\rangle \equiv \int [d A^\mu] U e^{-S_{YM}(A)} $. It is to be understood that 
$A_\mu \equiv A_\mu^a \lambda^a $ where $ A_\mu^a $ are the QCD gauge 
fields and $ \{ \lambda^a \} $ are the representation matrices for the
fermions in the fundamental representation of the gauge group SU(3).
In (\ref{two}) $ S_0 $ is a static fermion
propagator. The analysis of ref\cite{ef} begins with the introduction
of propagators for quarks in motion. They satisfy
\bea 
(\not\hspace{-1.25mm}{\it D} - m) \, S(x,y;A) &=& \delta^4(x-y)
\label{six} 
\eea
and may be expanded about the static propagator
\bea 
S(x,y;A) &=& S_0(x,y;A) +\int d^4z S_0(x,z;A)\bgamma \cdot \bD
S(z,y;A) \label{seven} 
\eea
where $ S_0 $ obeys 
\bea
(D_0\gamma^0 - m) \, S_0(x,y;A) &=&  \delta^4(x-y) \label{eight}
\eea
and is given explicitly by
\bea 
S_0(x,y;A) &=& -\imath \theta(x^0 - y^0) e^{-\imath m(x^0 - y^0)}
\frac{1+\gamma^0 }{2} P(x^0, y^0) \delta(\bx -\by) \label{nine} \\ &&
\qquad \qquad \qquad-\imath \theta(y^0 - x^0) e^{-\imath m(y^0 - x^0)}
\frac{1-\gamma^0 }{2} P(x^0, y^0) \delta(\bx -\by) \, . \nonumber 
\eea
To order $ m^{-2} $, with $ x^0 > y^0 $, the non-static propagator is
given by
\bea 
\lefteqn{ \left[1+\frac{1}{4m^2}(\bD^2-g\bsigma\cdot\bB)\right]
  S^{++}(x,y;A) =S_0^{++}(x,y;A) \label{ten}} \\ && +\int d^4 \omega
S_0^{++}(x,\omega;A) \left[ \frac{1}{2m}(\bD^2-g\bsigma\cdot\bB) +
  \frac{\imath g}{4m^2}(\delta_{ij}-\imath \epsilon_{ijk} \sigma^k)E^i
  D^j \right] S^{++}(\omega,y;A) \nonumber 
\eea
where the projections are, $ S^{++} \equiv \frac{1+\gamma^0}{2} \, S
\, \frac{1+\gamma^0}{2} ,\, \, S^{+-} \equiv \frac{1+\gamma^0}{2} \, S
\, \frac{1-\gamma^0}{2} $ , and so forth. To obtain leading relativistic corrections to the
static interaction, $\epsilon(r) $ of (\ref{three}), the static fermion propagators of 
(\ref{two}) are replaced by these nonstatic ones. Then
\bea
G_{I,c} &=& \langle tr_D tr_\sL (S^{++}_1 +S^{+-}_1+S^{-+}_1+S^{--}_1) \label{new1} \\
&& \qquad \qquad \times \, P(y_1, y_2) C^{-1}
(S^{++}_2 +S^{+-}_2+S^{-+}_2+S^{--}_2) C P(x_2, x_1) \rangle \nonumber \\
&\sim &- tr_D\left( \frac{1+\gamma^0}{2}\otimes \frac{1-\gamma^0}{2} \tilde{I}_{l,c}
+\frac{1-\gamma^0}{2}\otimes \frac{1-\gamma^0}{2} \tilde{I}_{s1}
+\frac{1+\gamma^0}{2}\otimes \frac{1+\gamma^0}{2} \tilde{I}_{s2} \right) \nonumber \\
&=& - tr_p (\tilde{I}_{l,c} + \tilde{I}_{s1} + \tilde{I}_{s2} ) = - tr_p (\tilde{I}_{l,c} + \tilde{I}_{s,c} )
\equiv - tr_p \, \tilde{G}_{p,c} \label{new2}
\eea
where $ I_{l,c}(I_{s,c}) $ is identified as the antiquark-charge-conjugated large(small) 2x2 Pauli component
and $ tr_p $ is the trace operator for Pauli matrices.
When the large component only is accounted for the combined static and $ O(m^{-2}) $ spin-orbit corrections are
\bea 
V &=& -\frac{1}{T} (\tilde{I}_{l,c} ) = \epsilon(r) \label{fourteen}\\ && +
\left(\frac{\bs_1\cdot\bL_1}{2m^2_1} - \frac{\bs_2\cdot\bL_2}{2m^2_2}
\right) \frac{\epsilon^\prime(r)}{r} \label{twelve} \\ &&
+\left(\frac{\bs_1\cdot\bL_1}{m^2_1} - \frac{\bs_2\cdot\bL_2}{m^2_2}
\right) \frac{V_1^\prime}{r} + \left(\frac{\bs_2\cdot\bL_1}{m_1 m_2} -
  \frac{\bs_1\cdot\bL_2}{m_1 m_2} \right) \frac{V_2^\prime}{r}
\label{thirdteen} \\ && - \left( \frac{\bs_1}{m_1} \cdot \frac{1}{T}
  \int^{T/2}_{-T/2} dz\langle \imath g \bB(\bx_1,z)\rangle_W /\langle
  1 \rangle_W - ( 1 \to 2 ) \right) \label{fifteen} 
\eea
where, $\bL_i \equiv \br \times \bp_i $ , and
\bea 
\left(\frac{\bs_1\cdot\bL_1}{m^2_1} -
  \frac{\bs_2\cdot\bL_2}{m^2_2} \right) \frac{V_1^\prime}{r} & \equiv
& - \bigg( \frac{\bs_1}{2m_1^2} \cdot \frac{1}{T} \int
\int^{T/2}_{-T/2} dz dz^\prime \langle \bB(\bx_1,z)
\bD^2(\bx_1,z^\prime) \rangle_W/\langle 1 \rangle_W \nonumber\\ &&
\qquad \qquad \qquad \qquad \qquad \qquad - (1 \to 2) \bigg)
\label{sixteen}\\ \left(\bs_2\cdot\bL_1 -\bs_1\cdot\bL_2 \right)
\frac{V_2^\prime}{r} & \equiv & - \bigg( \frac{\bs_1}{2} \cdot
\frac{1}{T} \int \int^{T/2}_{-T/2} dz dz^\prime \langle \bB(\bx_1,z)
\bD^2(\bx_2,z^\prime) \rangle_W/\langle 1 \rangle_W \nonumber \\ &&
\qquad \qquad \qquad \qquad \qquad \qquad - (1 \leftrightarrow 2)
\bigg) \, . \label{seventeen} 
\eea

A few remarks concerning this result are in order. First, the beginning four point function (\ref{one}) differs from that of ref
\cite{ef} where the antifermion fields have not been charge conjugated. This difference in four-point functions naturally yields a corresponding difference in the interactions derived from them. Second, in the interaction potential derivation of \cite{ef} there are several algebraic errors leading to the final result there (see appendix for details). The interaction resulting from the
algebraically correct non-charge-conjugated derivation we identify in this paper as the E.F. result, $ V_{EF} $ of the appendix. This E.F. potential
contrasts with the above potential most strikingly in that quark and antiquark contributions (for each term type) appear in the E.F. potential with identical algebraic sign.
In the potential above however,  derived in the appendix as $ V_{EF,c} $,
they appear with opposite sign as one would expect for a $q\bar{q}$ state.
For this reason the four-point function of
(\ref{one}) is seen to have a more acceptable interpretation as a $
q\bar{q} $ propagator than the beginning non-charge-conjugated four-point function of \cite{ef} leading to the E.F. result.

In ref\cite{gr} it is pointed out that insofar as the derived interaction is related to a v.e.v. in a Lorentz invariant theory the interaction itself
must also be invariant. In fact, $V$ as given in (\ref{fourteen}), is
not Lorentz invariant without a G.R.. It could be argued that omission of the small component $ \tilde{I}_{s,c} $ in the potential definition (\ref{fourteen}) has ruined the invariance since $ \tilde{I}_{s,c} $  appears in the invariant 4-point function (\ref{new2}) on an equal footing with the large component. That this is not the case is shown by simply replacing $\tilde{I}_{l,c} $ in (\ref{fourteen}) with $ \tilde{G}_{p,c} $
of (\ref{new2}) and then testing the resulting potential for invariance. The addition to V on this replacement exactly cancels line (\ref{twelve}), the classical spin-orbit and Thomas precession line, resulting in a spin-orbit interaction which we denote $V_{I,c} $.
Details of the derivation are given in the appendix, equations (\ref{50a})-(\ref{51a}). We now consider a Lorentz boost of $V_{I,c}$ to leading order in velocity. Care should be taken when transforming the
static propagators since by construction they must remain solutions to the non-covariant equation (\ref{eight}). To leading order their transformation is
\bea 
S_0^{\pm \pm}(x,y;A) &\to& S_0^{\pm \pm}(x,y;A) \pm \int d^4z
S_0^{\pm \pm}(x,z;A) \bv \cdot \bD(\bx, z) S_0^{\pm \pm}(z,y;A)
\label{eighteen} 
\eea
which is effectively carried out on their time-like path ordered
exponentials as 
\bea
P(x^0,y^0) &\to& P(x^0,y^0) - \imath
\int^{x^0}_{y^0} dz P(x^0,z)\bv \cdot \bD(\bx,z) P(z,y^0) \, .
\label{nineteen} 
\eea
And so the effective transformation (\ref{nineteen}), carried out in
(\ref{fifteen}), cancels with the momentum transformation of
(\ref{thirdteen}), and the remaining magnetic field transformation of (\ref{fifteen}) 
yields $ V_{I,c} \to V_{I,c} +
(\bs_1 \cdot [\br \times \bv]/m_1 -\bs_2 \cdot [\br \times \bv]/m_2)
\epsilon^\prime(r)/r $, and noninvariance.

What must be remembered about the v.e.v. $G_{I,c}$ of (\ref{new1}) is that its Lorentz invariance depends crucially upon its Dirac (or equivalently, its Pauli) trace operator without which it is not invariant
\bea
G_{I,c} &\to& G_{I,c} + \delta G \\
\delta G &\sim & tr_p \bigg( 
\b1_{2x2}\otimes \frac{\bsigma}{m_2} \cdot \int^{ \frac{T}{2} }_{-\frac{T}{2} }dz \lan \bE(\bx_2,z) \ran_W \times\bv \\ && \qquad \qquad + 
\frac{\bsigma}{m_1}\cdot \int^{\frac{T}{2} }_{-\frac{T}{2} }dz \lan \bE(\bx_1,z) \ran_W \times\bv \otimes \b1_{2x2}\bigg) \to 0 
\eea
(see appendix for details). From the potential definition (\ref{fourteen}) both as given or with $ \tilde{G}_{p,c} $ in place of the large component $\tilde{I}_{l,c} $ it is clear that Lorentz invariance of this potential related as it is to $ G_{I,c} $
is not to be expected, cannot be argued as it stands, i.e., without the  operation of the Pauli trace on $\tilde{I}_{l,c} $ ( or $ \tilde{G}_{p,c} $ ). In such an operation of course any spin dependence is averaged over, destroyed. Hence there remains no argument for the Lorentz invariance of the spin-dependent contribution to $V_{I,c}$ derived as it is from the invariant $G_{I,c}$, and so no argument for Gromes relations concerning $V_{I,c}$.

While $V_{I,c}$ like $V_{EF,c}$ is not Lorentz invariant, it does account consistently for the $O(m^{-2})$ contributions to the invariant 4-point function from which it is derived, while $V_{EF,c}$ does not. Considered as a $ Q\bar{Q}$ interaction Hamiltonian however $V_{I,c}$ has consistency problems of its own.
A simple example will suffice to make the point. It is well-known that in the heavy antiquark limit, $m_2 \to \infty $, lines (\ref{fourteen}) and (\ref{twelve}) agree exactly with the standard
$ m_1^{-2}$ expansion of the Dirac equation in Pauli form in the presence of a
central electric field. Then with the absence of line (\ref{twelve})
in $ V_{I,c}  $ there is disagreement between $ V_{I,c} $ and the
standard reduction of the Dirac equation. 

There is no such disagreement with the interaction derived from a four-point function whose Lorentz spinor structure is maintained (see appendix for details). This potential we denote $ V_{cov} $ since it is derived from the covariant four-point function.
Reduction of the covariant four-point function's off diagonal elements to $ O(m^{-2}) $ is performed by the Foldy-Wouthysen transformation,
$ U = \exp(\imath s(\xi))$, where, $ s(\xi) = \imath \gamma^0 \bgamma
\cdot \bD(\xi) /2m $ with a resulting interaction equivalent to $V$ of (\ref{fourteen}). I.e., the covariant interaction is equivalent to the invariant interaction when the small component contribution to the invariant trace is (improperly) omitted. Its
Lorentz transformation properties are unambiguous.  From the lowest
order transformation, $ \bp_i \to \bp_i -m_i \bv , \, \bB \to \bB -\bv
\times \bE $, and (\ref{nineteen}), $ V$ transforms as, $ V \to V +
(\bs_1 \cdot [\br \times \bv]/2m_1 -\bs_2 \cdot [\br \times \bv]/2m_2)
\epsilon^\prime(r)/r \equiv V + \tilde{V} $ (notice that neither $V_1
$ nor $V_2$ contribute to $ \tilde{V} $ ) in strict agreement with
its covariant transformation as a product of spinor products
\bea 
V \sim S_1^{++} \otimes C^{-1} S_2^{--} C &\to& L S_1^{++} L^{-1} \otimes
L C^{-1} S_2^{--}C L^{-1} \label{twentyone}\\ &\approx & (S_1^{++}
-\frac{1}{2}\bv\cdot\bgamma S_1^{-+} -\frac{1}{2} S_1^{+-}
\bv\cdot\bgamma )\otimes \\ && \qquad \qquad C^{-1} (S_2^{--}
+\frac{1}{2} S_2^{-+}\bv\cdot\bgamma + \frac{1}{2} \bv\cdot\bgamma S_2^{+-}
) C \\ &\sim & V +\tilde{V} 
\eea
where $ L \approx 1-\balpha \cdot \bv/2 $, again, as in the invariant case,  precluding the existence of Lorentz constraining relations between the transformed terms.

\section{ The Gromes method of derivation }

In view of the previous discussions a detailed examination of the G.R.
still seems worthwhile if only to answer to the enormous interest it
has generated over the last decade and a half. It should be clear however
that such an examination is little more than an academic exercise.
In ref\cite{gr} the
lowest order Lorentz transformation on $ V $ by which the relation is
derived is implemented in three distinct steps.

1)momentum transformation:
\bea
\qquad \qquad \qquad \qquad \bp_i \to \bp_i - m_i \bv \, . \nonumber
\eea

2)magnetic field transformation:
\bea
\qquad \qquad \qquad \qquad \bB \to \bB - \bv \times \bE \nonumber
\eea
\hspace{8mm} which by the given lemma, (2.8) of \cite{gr}, is
effectively carried out on the relative

\hspace{1.5mm} coordinate as
\bea 
\br \to \br +\bv \times \bs_1/m_1 - \bv \times \bs_2 /m_2 \, .
\nonumber 
\eea

3) $ \br_i $ transformation:
\bea 
\qquad \qquad \qquad \qquad \qquad \br_i \to \br_i + [\bv \times
\bs_i]/2m_i \nonumber 
\eea

\hspace{3.5mm} following the definition and transformation
specification of the quantity
\bea 
\bx &\equiv & \br_i +(\bp_i/m_i)t + [\bp_i \times \bs_i]/2m_i
\nonumber \\ \bx &\to & \bx - \bv t \, . \nonumber 
\eea

Effects of transformations 1) and 2) taken together with the effective
transformation of the path-ordered exponentials, (\ref{nineteen}),
combine in $ V $ of (\ref{fourteen}) to give the correct covariant
transformation (\ref{twentyone}), $ V \to V + \tilde{V} $.

Transformation 3) however is entirely spurious with no correspondence
to the field theoretic Lorentz transformation.  Its effect is to
produce the extra transformation terms, $ \epsilon(r) \to \epsilon(r)
+ ([\bv\times\bs_1]\cdot\br/2m_1
-[\bv\times\bs_2]\cdot\br/2m_2)\epsilon^\prime(r)/r $. The G.R.
appears when 1), 2), and 3) are taken together (with a sign error
included for the terms from step 2)) and invariance imposed to $ O
(m^{-1}) $. Then, $ \epsilon + V_1 - V_2 = 0 .  $

The relation has received derivation in several other contexts as
well. These derivations should each one undergo equal scrutiny.

\section{Summary}

The E.F. interaction Hamiltonian contains dynamical information on the
nonperturbative gluon field not present in models that simply
relativize the static linear potential. But in its given form one is
left either to make simplifying assumptions or to pursue evaluation on
the lattice.  For the latter, it is unclear whether the sign
discrepancies between the results here, (\ref{thirdteen}) -
(\ref{sixteen}), and ( $V_1, V_2 $ ) of ref\cite{ef} would affect
corresponding discrepancies between the respective lattice reductions.
For the former, the electric confinement ansatz, $V_1 = V_2 =0 $, in
conceptual agreement with Buchm\"{u}ller's picture\cite{bu}, assumed
in \cite{ef} and later abandoned (due to variance with spin
phenomenology), demonstrates that these should be made only with
special care and that they may in any event have consequences
difficult to predict. In these regards it may be viewed as a
shortcoming of the E.F. formalism that here too only the static limit
of the minimal area law enters explicitly; nonlocality is implied only
(e.g., via electric field insertions \cite{gr}). This happens as an
artifact of the initial fermion propagator expansion (\ref{seven})
around the static limit, and is readily remedied when these
propagators are replaced by ones more compatible with
semi-relativistic fermion motion. Then the Wilson loop of
(\ref{three}) is explicitly nonlocal and relates the resulting
potential directly to the minimum area.

Such a program has been carried through in an article by Brambilla,
Consoli, and Prosperi \cite{bram} both for spin and spin-independent
corrections to $ O(m^{-3}) $. From propagators expressed as integrals
over the phase space and the Nambu-Goto action as the effective area
their final potential is given in terms of familiar quantum mechanical
operators (It should be pointed out that the spin-orbit agreement with
ref\cite{gr} follows from a systematic mathematical error made in
their appendix \cite{ken}. See also ref\cite{thomas}. ).

A more rigorous implementation of the electric confinement ansatz
might also be attempted, though it is uncertain whether
Buchm\"{u}ller's picture remains self-consistent for quarks in motion
or whether it agrees with minimal area asymptotics. These
questions are considered in reference \cite{thomas}.

\vspace{10mm}

\hspace{-8mm} {\sl Acknowledgments}: This work was supported in part
by the National Science Foundation under Grant No. HRD-9154080. I am
also deeply indebted to The College of William \& Mary and in
particular to F. Gross for providing me office space and full access
to the Physics department facilities during the course of this work.

\section{appendix: potential derivations}

In evaluating the gauge-invariant $ Q\bar{Q} $ 4-point function
\bea
G &=& \langle 0| T^* \bar{\psi}(y_2)
P(y_2, y_1) \psi(y_1) \bar{\psi}(x_1) P(x_1, x_2) \psi(x_2) |0 \rangle \label{1a} \\
& \to & \langle tr
S(x_2, y_2 ;A) P(y_2, y_1) S(y_1, x_1 ;A) P(x_1, x_2) \rangle \label{2a}
\eea
the trace appearing in (\ref{2a}) may be taken over both gauge and Dirac matrices or over gauge matrices only, depending, respectively, on whether the four suppressed Dirac indices of (\ref{1a}) are of two like pairs or are all distinct. In the first instance the four point function is invariant under Lorentz transformations and in the second it is covariant, transforming as a product of spinor products. Interaction potentials derived from these invariant and covariant 4-point functions therefore have transformation properties that are invariant and covariant, respectively. Both 4-point functions appear in the literature as starting points for the potential derivation; for example, in
\cite{ef} the invariant version is used, and in \cite{bram} the covariant version.

\vspace{5mm}
{\bf Interaction potentials from the invariant 4-point function :}
\vspace{5mm}

First we derive interaction potentials from the invariant 4-point function.
The derivations follow closely that given in \cite{ef}.
Fermion propagators of (\ref{2a}) are expanded around the static propagator
as
\bea
S(x,y;A) &=& S_0(x,y;A) +\int d^4z S_0(x,z;A)\bgamma \cdot \bD
S(z,y;A)
\eea
with
\bea
S_0(x,y;A) &=& -\imath \theta(x^0 - y^0) e^{-\imath m(x^0 - y^0)}
\frac{1+\gamma^0 }{2} P(x^0, y^0) \delta(\bx -\by) \nonumber \\ &&
\qquad \qquad -\imath \theta(y^0 - x^0) e^{-\imath m(y^0 - x^0)}
\frac{1-\gamma^0 }{2} P(x^0, y^0) \delta(\bx -\by) \label{3a} \,.
\eea
They are then expanded over static energy projectors
\bea
S &=& \frac{1+\gamma^0}{2} \, S \, \frac{1+\gamma^0}{2} +
\frac{1+\gamma^0}{2} \, S \, \frac{1-\gamma^0}{2} +
\frac{1-\gamma^0}{2} \, S \, \frac{1+\gamma^0}{2}+
\frac{1-\gamma^0}{2} \, S \, \frac{1-\gamma^0}{2} \nonumber \\
&\equiv & S^{++} + S^{+-} + S^{-+} + S^{--}
\eea
so that
\bea
S^{++}(x,y;A) &=& S^{++}_0(x,y;A) +\int d^4z S^{++}_0(x,z;A)\bgamma \cdot \bD(z)
S^{-+}(z,y;A)\\
S^{+-}(x,y;A) &=& \int d^4z S^{++}_0(x,z;A)\bgamma \cdot \bD(z)
S^{--}(z,y;A)\\
S^{-+}(x,y;A) &=& \int d^4z S^{--}_0(x,z;A)\bgamma \cdot \bD(z)
S^{++}(z,y;A)\\
S^{--}(x,y;A) &=& S^{--}_0(x,y;A) +\int d^4z S^{--}_0(x,z;A)\bgamma \cdot \bD(z)
S^{+-}(z,y;A)
\eea
The projected propagator components above are evaluated to $O(m^{-2}) $ by substitution of the static propagator solution (\ref{3a}) and iteration to this order for $ x^0 > y^0 $. For the $ S^{++} $ component then 
\bea
S^{++}(x,y) &=& S^{++}_0(x,y) +\int d^4w \Delta^{+-}(x,w)\bgamma \cdot \bD(w)
S^{++}(w,y)
\eea
where
\bea
\Delta^{+-}(x,w) &=& \int d^4z S^{++}_0(x,z) \bgamma \cdot \bD(z)
S^{--}_0(z,w)  \nonumber \\ 
&=& - \frac{1+\gamma^0}{2} e^{-im(x^0+w^0)} \delta(\bx-\bw) \int dz_0 \theta(x^0-z^0) \theta(w^0-z^0) e^{2imz^0} f(z^0) \nonumber \\
&\equiv & - \frac{1+\gamma^0}{2} e^{-im(x^0+w^0)} \delta(\bx-\bw) I
\eea
and
\bea
f(z^0) &\equiv & P(x^0,z^0) \bgamma \cdot \bD(z) P(z^0,w^0) \, .
\eea
We rewrite the above integral as
\bea
I &=& e^{2imw^0}\theta(x^0-w^0) \int^0_{-\infty} dz^0 e^{2imz^0} f(z^0+w^0) \\
&& +e^{2imx^0} \theta(w^0-x^0) \int^0_{-\infty} dz^0 e^{2imz^0} f(z^0+x^0) \, . 
\eea
and use the $ O(m^{-2}) $ integral relation 
\bea
\int^0_{-\infty} dz^0 e^{2imz^0} f(z^0+\xi) & \simeq & -\frac{i }{2m} f(\xi)
+\frac{1}{4m^2} f^\prime(\xi)  \label{6a}
\eea
with
\bea
f(w^0) &=& P(x^0,w^0) \bgamma \cdot \bD(w) \, , \quad f^\prime(w^0) = -iP(x^0,w^0) [ D^0(w), \bgamma \cdot \bD(w)] \\
f(x^0) &=& \bgamma \cdot \bD(x)P(x^0,w^0) \, , \quad f^\prime(x^0) = -i 
[ D^0(x), \bgamma \cdot \bD(x)]P(x^0,w^0)  
\eea
to find on substitution
\bea
\Delta^{+-}(x,w) &=& -S_0^{++}(x,w) ( \frac{1}{2m} \bgamma \cdot \bD(w) +
\frac{1}{4m^2}[ D^0(w), \bgamma \cdot \bD(w)] ) \\ &&
 - ( \frac{1}{2m} \bgamma \cdot \bD(x) + \frac{1}{4m^2}[ D^0(x), \bgamma \cdot \bD(x)] ) S_0^{--}(x,w)
\eea
and
\bea
S^{++}(x,y) &=& S^{++}_0(x,y) - \int d^4w S_0^{++}(x,w) ( \frac{1}{2m} \bgamma \cdot \bD(w) \\ &&+ \frac{1}{4m^2}[ D^0(w), \bgamma \cdot \bD(w)] )
\bgamma \cdot \bD(w) S^{++}(w,y) \\ &&
- ( \frac{1}{2m} \bgamma \cdot \bD(x) + \frac{1}{4m^2}[ D^0(x), \bgamma \cdot \bD(x)] ) \tilde{\Delta }^{-+}(x,y)
\eea
where
\bea
\tilde{\Delta}^{-+}(x,y) &=& \int d^4w S^{--}_0(x,w) \bgamma \cdot \bD(w)
S^{++}(w,y) \\ 
&\simeq & - \frac{1}{2m} \bgamma \cdot \bD(x) S^{++}(x,y)\, .
\eea
With this, and from the identities
\bea
(\bgamma \cdot \bD)^2 &=& - \bD^2 +g\bsigma \cdot \bB \\
\left[ D^0, \bgamma \cdot \bD \right] \bgamma \cdot \bD  &=& -ig(\delta_{ij}-\imath \epsilon_{ijk} \sigma^k)E^i D^j
\eea
we arrive at
\bea
\lefteqn{ \left[1+\frac{1}{4m^2}(\bD^2-g\bsigma\cdot\bB)\right]
  S^{++}(x,y) =S_0^{++}(x,y) \label{5a}} \\ && +\int d^4 \omega
S_0^{++}(x,\omega) \left[ \frac{1}{2m}(\bD^2-g\bsigma\cdot\bB) +
  \frac{\imath g}{4m^2}(\delta_{ij}-\imath \epsilon_{ijk} \sigma^k)E^i
  D^j \right] S^{++}(\omega,y) \nonumber
\eea
equation (\ref{ten}). Then finally, to $ O(m^{-2}) $
\bea
S^{++}(x,y)&\simeq & S_0^{++}(x,y) +\int d^4 \omega
S_0^{++}(x,\omega) \bigg[ \frac{1}{2m}(\bD^2-g\bsigma \cdot \bB) \label{10a} \\ && \qquad \qquad \qquad +
  \frac{\imath g}{4m^2} (\delta_{ij}-\imath \epsilon_{ijk} \sigma^k)E^i
  D^j \bigg] S_0^{++}(\omega,y) \label{61a} \\
&&+\frac{1}{4m^2}\int d^4 \omega d^4 z S_0^{++}(x,\omega) (\bD^2-g\bsigma\cdot\bB) \\ && \qquad \qquad \qquad \times
S_0^{++}(\omega,z) (\bD^2-g\bsigma\cdot\bB)
S_0^{++}(z,y) \\
&& -\frac{1}{4m^2}(\bD^2-g\bsigma\cdot\bB) S_0^{++}(x,y) \, .
\eea
By similar methods we find ( $ x^0 > y^0 $ )
\bea
S^{+-}(x,y)&\simeq& - \frac{1}{2m} S_0^{++}(x,y) \bgamma \cdot \bD(y) \\
S^{-+}(x,y)&\simeq& - \frac{1}{2m} \bgamma \cdot \bD(x) S_0^{++}(x,y) \\ 
S^{--}(x,y) &\simeq & \frac{1}{4m^2} \bgamma \cdot \bD(x) S_0^{++}(x,y) 
\bgamma \cdot \bD(y) \\
&=& - \frac{1}{4m^2}\bigg[ \tilde{S}_0^{--}(x,y)[ \bD^2(y)-g\bsigma\cdot\bB(y)] \\ &&
+ig\int d^4z \tilde{S}_0^{--}(x,z) (\delta_{ij}-\imath \epsilon_{ijk} \sigma^k)E^i(z)
\tilde{S}_0^{--}(z,y)D^j(y)\bigg]
\eea
and
\bea
S^{--}(y,x)&\simeq & S_0^{--}(y,x) +\int d^4 \omega
S_0^{--}(y,\omega) \bigg[ \frac{1}{2m}(\bD^2-g\bsigma\cdot\bB) \\ &&
\qquad \qquad \qquad
  \qquad -\frac{\imath g}{4m^2} (\delta_{ij}-\imath \epsilon_{ijk} \sigma^k)E^i
  D^j \bigg] S_0^{--}(\omega,x) \label{62a} \\
&&+\frac{1}{4m^2}\int d^4 \omega d^4 z S_0^{--}(y,\omega) (\bD^2-g\bsigma\cdot\bB) \\ && \qquad \qquad \qquad \qquad \times
 S_0^{--}(\omega,z) (\bD^2-g\bsigma\cdot\bB) S_0^{--}(z,x) \\
&& -\frac{1}{4m^2}(\bD^2-g\bsigma\cdot\bB) S_0^{--}(y,x) \\
S^{-+}(y,x) &\simeq& -\frac{1}{2m} S_0^{--}(y,x) \bgamma \cdot \bD(x) \\ 
S^{+-}(y,x) &\simeq& - \frac{1}{2m} \bgamma \cdot \bD(y) S_0^{--}(y,x) \\
S^{++}(y,x) &\simeq & \frac{1}{4m^2} \bgamma \cdot \bD(y) S_0^{--}(y,x) 
\bgamma \cdot \bD(x) \\
&=& - \frac{1}{4m^2}\bigg[ \tilde{S}_0^{++}(y,x) [\bD^2(x)-g\bsigma \cdot\bB(x)] \\ &&
-ig\int d^4z \tilde{S}_0^{++}(y,z) (\delta_{ij}
-\imath \epsilon_{ijk} \sigma^k)E^i(z) \tilde{S}_0^{++}(z,x)D^j(x)\bigg] \label{11a}
\eea
where the identity
\bea
\bgamma \cdot \bD(x) S_0^{\pm \pm}(x,y)
&=& \tilde{S}_0^{\mp \mp}(x,y) \bgamma \cdot \bD(y)
\pm i \int d^4 z \tilde{S}_0^{\mp \mp}(x,z) \bgamma \cdot \bE(z) S_0^{\pm \pm }(z,y) \nonumber \\
&& \label{70a}
\eea
with
\bea
\tilde{S}_0(x,y;A) &\equiv & -\imath \theta(x^0 - y^0) e^{-\imath m(x^0 - y^0)}
\frac{1-\gamma^0 }{2} P(x^0, y^0) \delta(\bx -\by) \nonumber \\ &&
\qquad \qquad -\imath \theta(y^0 - x^0) e^{-\imath m(y^0 - x^0)}
\frac{1+\gamma^0 }{2} P(x^0, y^0) \delta(\bx -\by)
\eea
has been used, and the off-diagonals, $ S^{+-} $ and $ S^{-+} $ are shown to lowest order. Note that in evaluating $ S^{--}(y,x) $ above the integral relation used in place of equation (\ref{6a}) ( which is used in the evaluation of $ S^{++}(x,y) $ ) is
\bea
\int_0^{\infty} dz^0 e^{-2imz^0} f(z^0+\xi) & \simeq & -\frac{i }{2m} f(\xi)
- \frac{1}{4m^2} f^\prime(\xi)  \label{7a}
\eea
yielding the algebraic sign difference between lines (\ref{61a}) and (\ref{62a}).
For construction of the potential interaction we express the invariant Green's function
as
\bea
G_I &=& \langle tr \, S_1 \, P \, S_2 \, P \rangle \label{50a} \\
&=& tr_D \langle tr_\lambda \, S_1 \, P \, S_2 \, P \rangle \label{31a} \\
&\equiv& - tr_D
\tilde{G} e^{-i(m_1+m_2)(x^0-y^0)} \delta(\bx_1-\by_1)\delta(\bx_2-\by_2) \\
&=& - tr_p
\tilde{G}_p e^{-i(m_1+m_2)(x^0-y^0)} \delta(\bx_1-\by_1)\delta(\bx_2-\by_2)
\label{32a}
\eea
where in (\ref{31a}) Dirac and gauge traces have been explicitly separated, 
$ tr_p $ is the trace operator for Pauli matrices,
and $ \tilde{G}_p $ is the sum of 2x2 Pauli components along the diagonal
of  $ \tilde{G} $. The ``invariant'' potential interaction is then given by
\bea
V_I &=& -\frac{1}{T} \ln \tilde{G}_p \, . \label{33a}
\eea
for large $ T $. 
Substitution of (\ref{10a}) - (\ref{11a}) into (\ref{2a}) then yields for the invariant 4-point function 
\bea
G_I  &=& \langle tr_D tr_\lambda
[  S^{++}(x_1, y_1) + S^{--}(x_1, y_1) ]
P(y_1, y_2) \label{41a} \\ && \qquad \qquad \qquad \times
[ S^{--}(y_2, x_2) + S^{++}(y_2, x_2) ] P(x_2, x_1) \rangle
\\ &\equiv & - tr_D \langle tr_\lambda P \left[
\frac{1+\gamma^0}{2} S^+_{1p} + \frac{1-\gamma^0}{2} S^-_{1p} \right] 
 \otimes \left[ \frac{1-\gamma^0}{2} S^-_{2p} + \frac{1+\gamma^0}{2} S^+_{2p} \right]  \\ && \qquad \qquad \quad \times
\exp(\imath g \oint dz_\mu A^\mu(z)) \rangle
e^{-i(m_1+m_2) T} \delta(\bx_1-\by_1) \delta(\bx_2-\by_2) \label{9a}
\eea
where $ x^0 =-y^0 = \frac{T}{2} $ has been taken, and the ``p'' subscript identifies Pauli components which are given by
\bea
S^+_{1p} &=& 1 - \frac{i}{2m_1} \int^{ \frac{T}{2}}_{-\frac{T}{2}} dz
( \bD^2(\bx_1,z) - \bsigma_1 \cdot \bB(\bx_1,z) ) \label{30a} \\
&& - \frac{i}{4m_1^2} \int^{ \frac{T}{2}}_{-\frac{T}{2}} dz \,
\, \epsilon_{ijk} \sigma_1^k E^i(\bx_1,z)D^j(\bx_1,z) \\
S^-_{1p} &=& \frac{i}{4m_1^2} \int^{ \frac{T}{2}}_{-\frac{T}{2}} dz \,
\, \epsilon_{ijk} \sigma_1^k E^i(\bx_1,z)D^j(\bx_1,-T/2) \\
S^-_{2p} &=& 1 - \frac{i}{2m_2} \int^{ \frac{T}{2}}_{-\frac{T}{2}} dz
( \bD^2(\bx_2,z) - \bsigma_2 \cdot \bB(\bx_2,z) ) \\
&& +\frac{i}{4m_2^2} \int^{ \frac{T}{2}}_{-\frac{T}{2}} dz \,
 \epsilon_{ijk} \sigma_1^k E^i(\bx_2,z)D^j(\bx_2,z) \\
S^+_{2p} &=& - \frac{i}{4m_2^2} \int^{ \frac{T}{2}}_{-\frac{T}{2}} dz \,
\epsilon_{ijk} \sigma_2^k E^i(\bx_2,z)D^j(\bx_2, T/2)
\eea
for $O(m^{-2})$ static and spin-orbit contributions only. Gauge fields evaluated at $ \pm \frac{T}{2} $ have been set to zero. 

To show explicitly that the Lorentz invariance of $G_I$ depends upon the operation of its Pauli trace we consider the leading order boost $ L \simeq 1 - \balpha\cdot\bv/2 $ on $G_I$
\bea
G_I &\to & G_I +\delta G_I  =  \langle tr \, S_1 \, P \, S_2 \, P \rangle \\
&&+\frac{1}{2}\lan tr \bigg[S_0^{++}\otimes P 
\left([S^{-+}, \bv \cdot \bgamma ]-[S^{+-}, \bv \cdot \bgamma ] \right) P + \\
&&  \qquad \qquad \qquad \qquad P 
\left([S^{-+}, \bv \cdot \bgamma ]-[S^{+-}, \bv \cdot \bgamma ] \right) P \otimes S_0^{--} \bigg] \ran
\eea
giving on substitution for $ S^{\pm\mp} $ and the use of identity (\ref{70a})
\bea
\delta G &\sim & tr_p \bigg( 
\b1_{2x2}\otimes \frac{\bsigma}{m_2} \cdot \int^{ \frac{T}{2} }_{-\frac{T}{2} }dz \lan \bE(\bx_2,z) \ran_W  \times\bv \\ && \qquad \qquad - 
\frac{\bsigma}{m_1}\cdot \int^{\frac{T}{2} }_{-\frac{T}{2} }dz \lan \bE(\bx_1,z) \ran_W \times\bv \otimes \b1_{2x2}\bigg) \to 0 \, .
\eea
From (\ref{32a}) then
\bea
\tilde{G}_p &=& \langle (S^+_{1p}+ S^-_{1p})\otimes (S^-_{2p}+ S^+_{2p})W\rangle \\ &\simeq &  \tilde{I}_l +\tilde{I}_s \label{40a}
\eea
with
\bea
\tilde{I}_l &\equiv & \langle S^+_{1p} \otimes S^-_{2p}W\rangle \label{42a}
\label{65a}  \\
\tilde{I}_s &\equiv & \langle ( S^-_{1p}+ S^+_{2p})W \rangle
\eea
where $ \tilde{I}_{l} $ is the large component( identified by its nonzero
value in the static limit, $ m \to \infty $ ) and $ W = \exp(\imath g \oint dz_\mu A^\mu(z)) $ is the Wilson loop. From (\ref{33a}) we have for the interaction
\bea
V_I &=& -\frac{1}{T} \ln ( \tilde{I}_l +\tilde{I}_s ) \label{51a} \\
& \simeq & \epsilon(r) \\ &&
+\left(\frac{\bs_1\cdot\bL_1}{m^2_1} - \frac{\bs_2\cdot\bL_2}{m^2_2}
\right) \frac{V_1^\prime}{r} + \left(\frac{\bs_2\cdot\bL_1}{m_1 m_2} -
  \frac{\bs_1\cdot\bL_2}{m_1 m_2} \right) \frac{V_2^\prime}{r}
 \\ && - \left( \frac{\bs_1}{m_1} \cdot \frac{1}{T}
  \int^{T/2}_{-T/2} dz\langle \imath g \bB(\bx_1,z)\rangle_W /\langle
  1 \rangle_W + ( 1 \to 2 ) \right) 
\eea
where, $\bL_i \equiv \br \times \bp_i $ , and
\bea 
\left(\frac{\bs_1\cdot\bL_1}{m^2_1} -
  \frac{\bs_2\cdot\bL_2}{m^2_2} \right) \frac{V_1^\prime}{r} & \equiv
& - \bigg( \frac{\bs_1}{2m_1^2} \cdot \frac{1}{T} \int
\int^{T/2}_{-T/2} dz dz^\prime \langle \bB(\bx_1,z)
\bD^2(\bx_1,z^\prime) \rangle_W/\langle 1 \rangle_W \nonumber\\ &&
\qquad \qquad \qquad \qquad \qquad \qquad + (1 \to 2) \bigg)
\\ \left(\bs_2\cdot\bL_1 -\bs_1\cdot\bL_2 \right)
\frac{V_2^\prime}{r} & \equiv & - \bigg( \frac{\bs_1}{2} \cdot
\frac{1}{T} \int \int^{T/2}_{-T/2} dz dz^\prime \langle \bB(\bx_1,z)
\bD^2(\bx_2,z^\prime) \rangle_W/\langle 1 \rangle_W \nonumber \\ &&
\qquad \qquad \qquad \qquad \qquad \qquad + (1 \leftrightarrow 2)
\bigg) 
\eea
and the relation
\bea
\imath \int^{\frac{T}{2}}_{-\frac{T}{2}} dz \, \epsilon_{ijk} \sigma^k
\langle E^i(\bx,z) D^j(\bx,\xi) \rangle &\to & - T e^{-\epsilon(r) T} \, \,
\bsigma \cdot \bL \, \, \frac{\epsilon^\prime(r)}{r}
\eea
has been used. The sum $ \tilde{I}_l + \tilde{I}_s $ appearing above in (\ref{40a}) is
to be compared with $ \tilde{I} $ in equation (4.3) of \cite{ef}.
When the ``small'' component of the invariant trace, $ \tilde{I}_s $, is ignored $V_I$  becomes  
\bea
V_{EF}&=&  \epsilon(r) \label{8a} \\ && +
\left(\frac{\bs_1\cdot\bL_1}{2m^2_1} + \frac{\bs_2\cdot\bL_2}{2m^2_2}
\right) \frac{\epsilon^\prime(r)}{r} \label{60a} \\ &&
+\left(\frac{\bs_1\cdot\bL_1}{m^2_1} - \frac{\bs_2\cdot\bL_2}{m^2_2}
\right) \frac{V_1^\prime}{r} + \left(\frac{\bs_2\cdot\bL_1}{m_1 m_2} -
  \frac{\bs_1\cdot\bL_2}{m_1 m_2} \right) \frac{V_2^\prime}{r} 
 \\ && - \left( \frac{\bs_1}{m_1} \cdot \frac{1}{T}
  \int^{T/2}_{-T/2} dz\langle \imath g \bB(\bx_1,z)\rangle_W /\langle
  1 \rangle_W + ( 1 \to 2 ) \right) \label{15a}
\eea
with
\bea 
\left(\frac{\bs_1\cdot\bL_1}{m^2_1} -
  \frac{\bs_2\cdot\bL_2}{m^2_2} \right) \frac{V_1^\prime}{r} & \equiv
& - \bigg( \frac{\bs_1}{2m_1^2} \cdot \frac{1}{T} \int
\int^{T/2}_{-T/2} dz dz^\prime \langle \bB(\bx_1,z)
\bD^2(\bx_1,z^\prime) \rangle_W/\langle 1 \rangle_W \nonumber\\ &&
\qquad \qquad \qquad \qquad \qquad \qquad + (1 \to 2) \bigg) \label{16a}
\\ \left(\bs_2\cdot\bL_1 -\bs_1\cdot\bL_2 \right)
\frac{V_2^\prime}{r} & \equiv & - \bigg( \frac{\bs_1}{2} \cdot
\frac{1}{T} \int \int^{T/2}_{-T/2} dz dz^\prime \langle \bB(\bx_1,z)
\bD^2(\bx_2,z^\prime) \rangle_W/\langle 1 \rangle_W \nonumber \\ &&
\qquad \qquad \qquad \qquad \qquad \qquad + (1 \leftrightarrow 2)
\bigg) \label{17a} \, . 
\eea

The ``large component only'' interaction is precisely what is derived by authors Eichten and Feinberg in \cite{ef}. It's differences with the above $ V_{EF} $ potential are easily traced to three algebraic errors made in their derivation: 1) In their appendix there is an expression for $ S^{++}$, (A3), given in terms of an integral whose evaluation is given in (A15). Direct substitution of (A15) into (A3) does not yield their equation (2.11) for $ S^{++} $, but yields instead
equation (\ref{5a}) above with spin terms on the rhs differing in algebraic sign. 2) There is a misrepresentation of the coordinate space momentum operator in going from equation (4.11a) to (4.11b) in the E.F. work. 3) There is a missed sign in the evaluation of an integral for $ S^{--}(y,x) $ whose analogue in the evaluation of $ S^{++}(x,y) $ is (A7b). Here this mistake amounts to using the integral result of (\ref{6a})
in the evaluation of $ S^{--}(y,x) $ where the result of (\ref{7a}) is called for.
These individual errors would have the following effects on the above $ V_{EF} $: i)   error 1) would change the overall algebraic signs of lines (\ref{60a}) and(\ref{15a}). 
ii)  error 2) would partially correct error 1) by again changing the algebraic sign of line (\ref{60a}).
iii) error 3) would change the relative sign between terms in line (\ref{60a})
making the subscripted ``2'' term negative.
The resulting interaction Hamiltonian would then be identical with
the static plus spin-orbit interaction of reference \cite{ef}. 

The conspicuous lack of relative sign differences between subscripted ``1'' and ``2'' contributions in $V_{EF}$ above makes its interpretation as a $ Q\bar{Q}$
interaction Hamiltonian problematic. When the antiquark field operators in the beginning four point function (\ref{1a}) are properly charge conjugated ( as they are for example in \cite{bram})this problem disappears.
Specifically, the necessary change is
\bea
\psi(x_2) & \to & \psi^c(x_2) \\
\Rightarrow  S(x_2, y_2 ;A)  & \to & C^{-1} S(x_2, y_2 ;A) C = 
- \left[S(y_2, x_2 ;-A^T)\right]^T
\eea
where the outer transposition in the last step is taken over both gauge and Dirac matrices. The effect on $ V_{EF} $ of (\ref{8a}) from the above charge conjugation made in (\ref{2a}) is simply to change the algebraic signs of the field insertions on antiquark lines of the Wilson loop while preserving the original path ordering
\bea
V_{EF,c}&=&  \epsilon(r) \label{20a} \\ && +
\left(\frac{\bs_1\cdot\bL_1}{2m^2_1} - \frac{\bs_2\cdot\bL_2}{2m^2_2}
\right) \frac{\epsilon^\prime(r)}{r} \\ &&
+\left(\frac{\bs_1\cdot\bL_1}{m^2_1} - \frac{\bs_2\cdot\bL_2}{m^2_2}
\right) \frac{V_1^\prime}{r} + \left(\frac{\bs_2\cdot\bL_1}{m_1 m_2} -
  \frac{\bs_1\cdot\bL_2}{m_1 m_2} \right) \frac{V_2^\prime}{r}
 \\ && - \left( \frac{\bs_1}{m_1} \cdot \frac{1}{T}
  \int^{T/2}_{-T/2} dz\langle \imath g \bB(\bx_1,z)\rangle_W /\langle
  1 \rangle_W - ( 1 \to 2 ) \right) 
\eea
with
\bea 
\left(\frac{\bs_1\cdot\bL_1}{m^2_1} -
  \frac{\bs_2\cdot\bL_2}{m^2_2} \right) \frac{V_1^\prime}{r} & \equiv
& - \bigg( \frac{\bs_1}{2m_1^2} \cdot \frac{1}{T} \int
\int^{T/2}_{-T/2} dz dz^\prime \langle \bB(\bx_1,z)
\bD^2(\bx_1,z^\prime) \rangle_W/\langle 1 \rangle_W \nonumber\\ &&
\qquad \qquad \qquad \qquad \qquad \qquad - (1 \to 2) \bigg)
\\ \left(\bs_2\cdot\bL_1 -\bs_1\cdot\bL_2 \right)
\frac{V_2^\prime}{r} & \equiv & - \bigg( \frac{\bs_1}{2} \cdot
\frac{1}{T} \int \int^{T/2}_{-T/2} dz dz^\prime \langle \bB(\bx_1,z)
\bD^2(\bx_2,z^\prime) \rangle_W/\langle 1 \rangle_W \nonumber \\ &&
\qquad \qquad \qquad \qquad \qquad \qquad - (1 \leftrightarrow 2)
\bigg) \, . 
\eea
where the ``c'' subscript indicates that antiquark fields have been charge conjugated. This is our $ V $ of equation(\ref{fourteen}).

To obtain the full $V_I$ charge conjugated interaction likewise
simply requires that we change the algebraic signs of the field insertions 
along the antiquark lines appearing in $ V_I $ above
\bea
V_{I,c} &=& \epsilon(r) \\ && 
+\left(\frac{\bs_1\cdot\bL_1}{m^2_1} - \frac{\bs_2\cdot\bL_2}{m^2_2}
\right) \frac{V_1^\prime}{r} + \left(\frac{\bs_2\cdot\bL_1}{m_1 m_2} -
  \frac{\bs_1\cdot\bL_2}{m_1 m_2} \right) \frac{V_2^\prime}{r}
 \\ && - \left( \frac{\bs_1}{m_1} \cdot \frac{1}{T}
  \int^{T/2}_{-T/2} dz\langle \imath g \bB(\bx_1,z)\rangle_W /\langle
  1 \rangle_W - ( 1 \to 2 ) \right) 
\eea
with
\bea 
\left(\frac{\bs_1\cdot\bL_1}{m^2_1} -
  \frac{\bs_2\cdot\bL_2}{m^2_2} \right) \frac{V_1^\prime}{r} & \equiv
& - \bigg( \frac{\bs_1}{2m_1^2} \cdot \frac{1}{T} \int
\int^{T/2}_{-T/2} dz dz^\prime \langle \bB(\bx_1,z)
\bD^2(\bx_1,z^\prime) \rangle_W/\langle 1 \rangle_W \nonumber\\ &&
\qquad \qquad \qquad \qquad \qquad \qquad - (1 \to 2) \bigg)
\\ \left(\bs_2\cdot\bL_1 -\bs_1\cdot\bL_2 \right)
\frac{V_2^\prime}{r} & \equiv & - \bigg( \frac{\bs_1}{2} \cdot
\frac{1}{T} \int \int^{T/2}_{-T/2} dz dz^\prime \langle \bB(\bx_1,z)
\bD^2(\bx_2,z^\prime) \rangle_W/\langle 1 \rangle_W \nonumber \\ &&
\qquad \qquad \qquad \qquad \qquad \qquad - (1 \leftrightarrow 2)
\bigg) \, . 
\eea

\vspace{5mm}
{\bf Interaction potential from the covariant 4-point function :}
\vspace{5mm}

Here we derive the interaction potential from the covariant 4-point function
whose antiquark fields are charge conjugated.
\bea
G_{cov}  &=& \langle tr_\lambda 
S(x_1, y_1; A ) P(y_1, y_2) C^{-1} S(y_2, x_2;A) C P(x_2, x_1) \rangle \\
&\equiv & - \tilde{G} e^{-i(m_1+m_2) T} \delta(\bx_1-\by_1) \delta(\bx_2-\by_2) \, .
\eea
The interaction is found from the large component of the
diagonalized $ \tilde{G} $ by
\bea
V_{cov,c} &=& -\frac{1}{T} \ln \tilde{G}_l
\eea
where $ \tilde{G}_l $ is the large component.
The diagonalization to $ O(m^{-2} )$ is performed via the Foldy-Wouthysen transformation $ U = \exp(\imath s(\xi))$, with, $ s(\xi) = \imath \gamma^0 \bgamma \cdot \bD(\xi) /2m $. It will suffice to reduce the off diagonal elements to $ O(m^{-2}) $, accomplished by a single transformation.
For the quark propagator reduction, using the above results 
\bea
S^{++}(x,y) &=& S^{++}_0(x,y) + \int d^4w S_0^{++}(x,w) \frac{1}{2m} ( \bgamma \cdot \bD(w))^2  S^{++}_0(w,y) 
\\ && + O(m^{-2}) \\
S^{+-}(x,y)&=& - \frac{1}{2m} S_0^{++}(x,y) \bgamma \cdot \bD(y) + O(m^{-2}) \\
S^{-+}(x,y)&=& - \frac{1}{2m} \bgamma \cdot \bD(x) S_0^{++}(x,y) + O(m^{-2}) \\
S^{--}(x,y)& \sim & O(m^{-2})
\eea 
we have
\bea
S(x,y)& \to& S^\prime(x,y)  =  e^{iu(x)} S(x,y) e^{-iu(y)} \\
&=& S(x,y) +i[ u(x)(S^{++}(x,y)+ S^{+-}(x,y)+ S^{-+}(x,y)) \\ &&
\qquad \qquad -(S^{++}(x,y)+ S^{+-}(x,y)+ S^{-+}(x,y))u(y) \\ &&
-\frac{1}{2}[u(x)(u(x) S^{++}(x,y)-S^{++}(x,y) u(y)) \\ && 
\qquad \qquad -(u(x) S^{++}(x,y)-S^{++}(x,y) u(y)) u(y) ] +O(m^{-3}) \\
&=& S^{++}(x,y)+\frac{1}{8m^2}[( \bgamma \cdot \bD(x))^2  S^{++}_0(x,y)
+ S^{++}_0(x,y)( \bgamma \cdot \bD(y))^2 ] \nonumber \\
&& + \frac{1}{4m^2} \int d^4w [ S_0^{++}(x,w) ( \bgamma \cdot \bD(w))^2  S^{++}_0(w,y) \bgamma \cdot \bD(y) \\ && \qquad \quad
- \bgamma \cdot \bD(x) S_0^{++}(x,w) ( \bgamma \cdot \bD(w))^2  S^{++}_0(w,y)] +O(m^{-3})
\eea
The off diagonals are now of $ O(m^{-2})$. Another transformation would 
further reduce the off diagonals to  $ O(m^{-3})$ while contributing nothing
to the diagonal. The diagonalized quark large component is therefore given by $ S^{++}(x,y) $ alone ( with gauge fields again
set to zero at $ \pm \frac{T}{2} $ ). The antiquark propagator diagonalization
follows along the same lines, leading to the large antiquark component $ C^{-1} S^{--}(y,x)C $. Then from (\ref{41a}) to (\ref{42a})  above we find
\bea
\tilde{G}_l &=& \tilde{I}_{l,c }
\eea
where the ``c'' subscript indicates that field insertions on the antiquark
line of the Wilson loop in $ \tilde{I}_l $ of (\ref{65a}) have undergone an algebraic sign change. This gives
\bea
V_{cov,c} &=& V_{EF,c} \\ &=& V
\eea
where $ V_{EF,c} $ is given in (\ref{20a}) and $ V $ in (\ref{fourteen}). 
I.e. the interaction obtained from the diagonalized covariant 4-point function is identical with the one obtained from the large component of the invariant interaction.

\end{document}